\documentclass[twocolumn,aps,prl]{revtex4} % ketoszlopos
\usepackage{graphics}
\usepackage{epsfig}
\usepackage{graphicx}
\usepackage{indentfirst}
\newcommand{\kepnagysag}{7cm}
\begin{document}
\title{
Self-generated Self-similar Traffic
}
\author{
P{\'e}ter H\'aga, P{\'e}ter Pollner, G\'abor Simon, Istv\'an Csabai,  
G\'abor Vattay}
\affiliation{
Communication Networks Laboratory\\
E{\"o}tv{\"o}s Lor\'and University, H-1117 Budapest, P\'azm\'any P.s. 1/A.\\ 
\mbox{E-mail: \{haga, pollner, gaba, csabai, vattay\}@complex.elte.hu}
}
%\date{\today}
\begin{abstract}Self-similarity in the network traffic has been studied from
  several aspects: both at the user side and at the network side there
  are many sources of the long range dependence. Recently some
  dynamical origins are also identified: the TCP adaptive congestion
  avoidance algorithm itself can produce chaotic and long range
  dependent throughput behavior, if the loss rate is very high. In
  this paper we show that there is a close connection between the
  static and dynamic origins of self-similarity: parallel TCPs can
  generate the self-similarity themselves, they can introduce heavily
  fluctuations into the background traffic and produce high effective
  loss rate causing a long range dependent TCP flow, however, the
  dropped packet ratio is low.
\end{abstract}
\maketitle

\section{Introduction}

In a large number of internet traffic measurements many authors
detected self-similarity\cite{csabai,crovella,feldmann,leland}.
Self-similarity is usually attributed with heavy-tailed distributions
of objects at the traffic sources, e.g.  sizes of transfered
files\cite{leland,park} or time delays in user interactions\cite{crovella}.

Recently the dynamical origins of self-similarity also attract
increasing attention. Chaotic maps are known to generate fractal
properties\cite{erramilli}, and it has been shown, that the
exponential backoff algorithm of the TCP protocol can produce long
range dependent traffic under some special circumstances.
Veres\cite{veresboda} emphasized the chaotic nature of the transport,
and Fekete\cite{fekete} gave an analytical explanation that this
chaoticity is in close relation with the loss rate at highly utilized
buffers due to packet drops. Guo\cite{guo} gave a Markovian analysis of the
backoff mechanism, and showed, that the self-similarity can be
observed only at very high loss rates.

In this paper we study a lossless network, where the traffic sources
are {\em not} heavy-tailed distributed. Even though we observe self-similar
traffic, which is generated by the communicating TCPs themselves. We
show, that the random switching between the destinations of the flows
together with the complex dynamics of interacting TCPs
can lead to long range dependency. The interaction between TCPs at the
buffers leads to backoff phases in the individual TCPs causing fractal
traffic in an environment where the loss rate is much below the lower
bound of self-similarity.

The outline of the paper is the following: First we introduce the
concept of real and effective loss. Then we present a simple lossless
model network, where self-similar traffic is observed, however the
necessary conditions discusssed in the literature cited above, 
are not satisfied. Next we show that similar scenario
can be found in real networks as well. Finally we conclude our results.

%- self-similarity, long range dependence: general definitions and concepts

%- known sources of self similarity:\\
%Willinger: file size distribution\\
%Floyd: user interaction\\
%Crovella: high loss rate\\
%Veres: chaos due to small buffers\\
%Our: low real loss rate but high effective loss (solves the problem of
%Crovella's argument: high loss rate can emerge at good
%network-hardware conditions.)

\section{Real and effective loss}

In the internet traffic many individuals use a common, finite resource
to transmit information. If the resources are exhausted (e.g. routers
are congested), data throughput is not possible. Therefore data
transmitters should avoid congestion on shared information routes.
Most of today's computer programs use similar algorithm to avoid
congestion: they apply basicly the same TCP protocol with slight
differences. The common concept in every TCP is, that the data sending
rate must be adapted to the actually available
resources\cite{jacobson88,jacobson90}. Every TCP starts with a blind
measuring phase (slow start), which exponentially reaches the maximum
throughput rate. If the route, where TCP sends its data is stationary
utilized, the algorithm works in a high throughput -- slow adaption
phase (congestion avoidance). The sending rate is varied around a high
value by slowly increasing and rapidly decreasing it. Since every sent
packet and its received acknowledgement continuously measure the
transmission possibilities, this phase is very stable and can adapt to
slowly varying situations\cite{veresvattay}.

If the route gets highly loaded, the TCP tries to clear the congestion
by decreasing the sending rate. If the congestion is so high, that the
TCP cannot guess the proper sending rate (acknowledgements do not
arrive in the timeout period), the algorithm enters a very slow sending
phase (exponential backoff). In this phase due to the lack of
information an exponentially slowing algorithm is applied to try to find
a new possible sending rate: the last package is resent after
exponential increasing time intervals until an acknowledgement received
or a maximum time interval is reached.

In this paper we concentrate on the backoff phase of the TCP. We
will show, that due to its blind nature, in this phase the TCP can
feel higher loss rates as it really is.

By the blindness of the TCP we mean the consequence of Karn's
algorithm\cite{karn}, which governs the backoff phase. Under normal
transmission conditions TCP operates in slow start or in congestion
avoidance mode.  In these modes the TCP estimates the optimal sending
rate from the arrival time of the acknowledgements (ACK) by
calculating the average round trip time (SRTT) and its average
deviation from the mean (MDEV).  After each received ACK the TCP
estimates the retransmission timeout (RTO). If this timeout is
exceeded between sending a packet and receiving an acknowledgement for
it, the TCP retransmits the packet assumed to be lost (by real loss or
timeout). In this situation TCP applies the so called Karn's algorithm.

The Karn's algorithm specifies that the acknowledgments for
retransmitted data packets cannot be used to approximate the sending
rate. Since for a received ACK packet one cannot decide if it is the
ACK of the original or of the retransmitted packet, the round trip
time (RTT) and so the sending rate cannot be estimated. The RTT can
be calculated only for those packets, which are not retransmitted.  So
the TCP retransmits the packet and doubles the RTO calculated from the
previous flow-informations (backoff phase). If the retransmitted
packet timeouts again, the RTO is doubled and the packet is
retransmitted again. The RTO is increased up to a maximal value defined
in the protocol.  The TCP leaves the backoff phase only if the RTT can
be estimated without ambiguity: the TCP must receive the
acknowledgements of two consecutive sent packets.  We will show a
situation where this method reports reasonably higher loss rate for
the TCP as it really is.

We distinguish the loss into real or virtual. {\em Real} loss is referred to
dropped packets which either are not arrived to the destination or the
acknowledgment for it do not arrive to the sending TCP. We call a loss
to be virtual if the acknowledgment arrives after the
retransmission timeout (RTO) period, so the packet is retransmitted due
to a spurious timeout. The {\em effictive} loss is assembled from
the real and virtual losses.

This distinction is important, since real loss emerges at highly
congested buffers or at low quality lines (e.g. radio connections).
These situations can be solved by improving the hardware conditions.
In contrast, high virtual loss can evolve also under very good
hardware conditions from heavily fluctuating background traffic. On
a route with several routers, where the packets can stay
in a long queue, round
trip times can change in a wide range depending on the queuing time.
The queuing time depends on the saturation of the buffers on the route.
If the background traffic fills the buffers at a varying rate, the
queueing time, and so the round trip time varies also. Bursty
background traffic can fill the buffers rapidly to a high value, and
after that can leave it to be cleared out.

If the round trip time increases to such a high value due to a rapid
filling up, that it becomes larger than the retransmission timeout
value, a virtual loss occurs. After a burst which caused the virtual
loss the clearing out of the buffer will lead to a shorter round 
trip time, which
decreases the RTO value also. So for the next burst event the RTO is
not large enough that the TCP can receive the ACK packet. So another
virtual loss occurs without really loosing the sent packets.

We will show in a model network and in a real measurement, that long
range dependent traffic can emerge from the virtual losses due to the
bursty background, however, real packet loss rate is so low, that one
would expect a scalable simple traffic rate.

%- TCP congestion avoidance algorithm
%Karn's Algorithm in W. Richard Stevens TCP/IP Illustrated
%
%\cite{karn87} specify that when a timeout
%and retransmission occur, we cannot update the RTT estimators
%when the acknowledgment for the retransmitted data finally arrives.
%This is because we don't know to which transmission the ACK corresponds.
%(Perhaps the first transmission was delayed and not thrown away,
%or perhaps the ACK of the first transmission was delayed.)
%
%Also, since the data was retransmitted, and the exponential
%backoff has been applied to the RTO, we reuse this backed off
%RTO for the next transmission. Don't calculate a new RTO until
%an acknowledgment is received for a segment that was not
%retransmitted.
%
%- real loss and virtual loss
%
%- bursts in the background traffic: cause of virtual loss

\section{Case study: simulation}

%- ns simulation with fixed file sizes

In this section we present a simple model network, which shows
self-similar traffic. Our model differs in several aspects from
previous studies in the literature excluding the known reasons of
self-similarity. 

\begin{figure}[h]
\centerline{\resizebox{\kepnagysag}{!}{\includegraphics{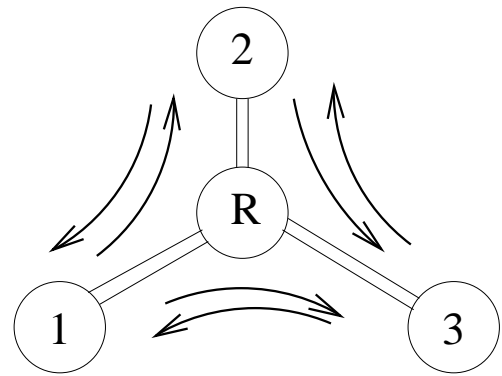}}}
\caption{The topology of the network model. The numbered circles represent
hosts with TCP agents, and $R$ denotes the router.\label{fig-topol}}
\end{figure}

In our model three hosts transfer fixed sized files to each other
through a router. All hosts transfer files with the same size.  The
topology of the model is shown in Fig.~\ref{fig-topol}.  From each
numbered sites of the network a TCP flow is initiated to one of the
other numbered sites. Each TCP flow passes through the router $R$
using full duplex connections, so the flow of the acknowledgements 
do not interfere with the corresponding TCP data flow. However data from other
TCPs must share the same buffers and lines with acknowledgements.

We have chosen the network to be practically lossless: the buffer
length in the router was set so large, that it is very improbable that
TCP flows fill them.  All the six buffers for the three full duplex
lines are large enough to store all the files transfered between the
hosts at a moment.  There is no external packet loss on the lines as
well.

We will study the traffic properties on a {\em line} connecting a
chosen host with the router. So the packet flows we want to analyze
are initiated from a fixed chosen host and they are built up from
several successive TCP flows. 

\begin{figure}[h]
\centerline{\resizebox{\kepnagysag}{!}{\includegraphics{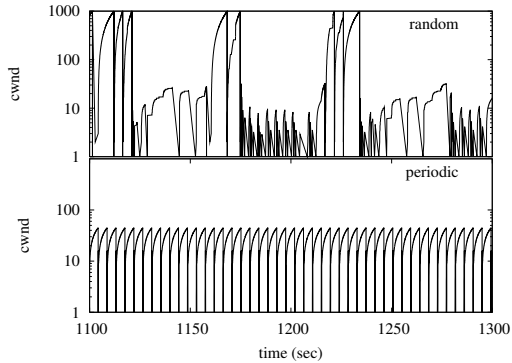}}}
\caption{The congestion window for one of the hosts. a) regular case
b) random case \label{fig-throughput}}
\end{figure}

In this topology the traffic is not always self-similar. The
throughput of packets on a line can be regular if the destination of
the individual TCP flows is chosen on a regular way.  An example is
shown in Fig.~\ref{fig-throughput}a, where the TCP flows has been
generated with the following simple rule: from the host numbered by
$i$ ($i=1\ldots 3$) the TCP sends packets to host $(i\ {\rm mod}\ 3) +1$.
After a file has been transmitted, the host starts a new TCP flow {\em
  immediately}, there is no external random delay between the flows as
it would be if we took the user behavior into
account\cite{danzig,paxsonfloyd}. Under such regular sending rules the
TCPs can utilize the available bandwidth and the traffic has a
scalable periodic pattern. In Fig.~\ref{fig-throughput}a we show the
congestion window for a host.

We have implemented this simple network in the ns-2 simulator (version number: 2.\ 1b7a)\cite{netsim}.
The link parameters are: link rate 1Mbps, delay 1ms. File size of a
TCP flow was 1000 packet. The receiver window was much larger than the
file sizes. We have used the Reno version of TCP.

If we introduce stochasticity in the sending rules we arrive at a
non-scalable, long range dependent traffic. We applied the following
rules to generate self-similar packet transport. All hosts send fixed
size files to each other. Each host uses the same file size.  If a
host starts a new TCP, it randomly chooses to which host to send the
file. After a transmission is completed, the host chooses the next
destination immediately. The next destination is chosen randomly again
without silent period between consecutive TCP flows.  In
Fig.~\ref{fig-throughput}b we show, that the stochasticity inhibits
the TCPs to synchronize and the packet transport becomes irregular.
The size of the transfered files was chosen not too large to hinder
the TCPs to adapt to each other.

We investigate now the irregular packet traffic if it shows
self-similarity or not. 
%Here we speak about self-similarity however
%such traffic should be classified in a very accurate language as
%pseudo-self-similar, since the self-similarity is valid only on a
%finite time scale due to the upper bounds for timeout variables in the
%TCP protocols[Crovella].  Even though we use the shorter form, since in
%numerical models and measurements the word self-similar is used if the
%property holds at least for 2-3 decades[Vicsek]. In exact theoretical
%models the self-similarity should be used only if scaling is valid for
%infinite range of course.
%
Self-similarity can be tested by investigating the second order
statistics of the traffic\cite{leland}. Consider a weakly stationary
process $X$, with constant mean and autocorrelation function $c(k)$.
Let $X^{m}(k)=\frac{1}{m}\sum_{i=(k-1)m+1}^{km} X(i)$ denote the $m$
aggregated series of $X$. The process $X$ is self-similar if $X\approx
m^{1-H}X^{m}$, and $X$ is second order self-similar if $m^{1-H}X^{m}$
has the same variance and autocorrelation as $X$. The sign $\approx$
expresses the fact that the equality can be satisfied only in a
stochastic sense, exact equation can only be used for abstract
mathematical objects.

\begin{figure}[h]
\centerline{\resizebox{\kepnagysag}{!}{\includegraphics{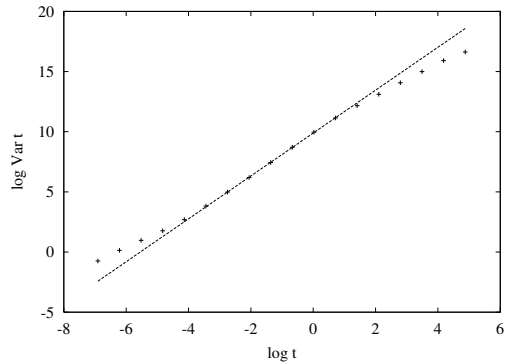}}}
\caption{Variance time plot of the aggregated traffic on a host-router link.
The fitted line indicates $H=0.89$.
\label{fig-aggtimemod}}
\end{figure}

We have performed self-similarity test by using the variance time
method.  In Fig.~\ref{fig-aggtimemod} we plot the variance of the
aggregated time series of the packets which scales as
$$
Var(\delta X)=(X^{m}(k)-\langle X^{m}\rangle_k )^2 \sim m^{2H}.
$$
The fitted line in the figure indicates Hurst exponent $H=0.89$ showing that the
time series is self-similar since $H>0.5$. We emphasize again, that 
the time series under consideration is built up from several consecutive
TCP flows.

%subsec

If a traffic is self-similar it shows properties which differs from
ones of memory-less (Markovian) processes: the dynamics of the system
shows a strong history dependence. This long range dependence in the
time evolution manifests itself typically in heavy-tailed
distributions. A distribution is heavy-tailed if asymptotic decay of
the distribution is slow: a power-law with exponent less than two. Due
to the always existing upper bounds of the measured data it is enough
if the decay holds in the last decades below the maximum value:
$$
P(S>s) \sim s^{-(1+\alpha)}, as\ \ s\in [10^{-n}s_{max},s_{max}],\ {\rm and} 
\ n>2.5\ .
$$
Such distributions are called heavy-tailed, since occurrence of
values much larger than the mean value of the observable $s$ is not
negligible in contrast to commonly used distributions as Gaussian or
exponential. However in measured time series it can happen, that from
the tail we cannot detect so many events as it is needed to plot a
smooth distribution function. In these cases it is favorably to work
with the {\em cumulative} distribution $P_c(s)=\int^s P(s')ds'$, which
has an asymptotic behavior as $P_c(s)\sim 1-s^{-\alpha}$. Therefore
one should use the inverse cumulative function $1-P_c(s)$ to fit the
parameter $\alpha$ on the logarithmic plot.

\begin{figure}[h]
\centerline{\resizebox{\kepnagysag}{!}{\includegraphics{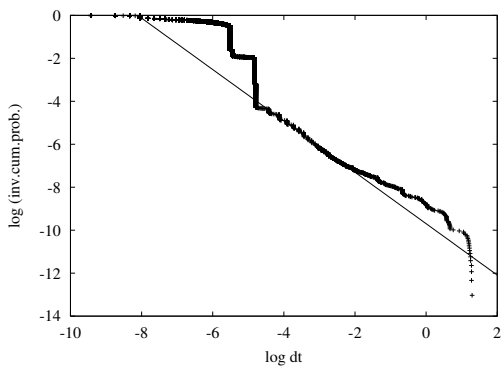}}}
\caption{The distribution of the packet inter arrival times.
The fitted line indicates $H=0.91$.
\label{fig-decaymod}}
\end{figure}

Now we want to investigate if the long range dependency shows up in
the traffic. We consider only the case when a destinations of the TCPs
were chosen randomly. In Fig.~\ref{fig-decaymod} we plot the inverse cumulative
distribution of the packet inter arrival time on a down link.  The
distribution shows a slow decay with $\alpha=1.18$ which indicates that the
fluctuating traffic has long silent periods. A power law decaying
fluctuation can be characterized by the Hurst exponent if the traffic
is viewed as an ON-OFF process. The silent periods are the OFF
intervals of the process. The Hurst parameter is known for this type
of process\cite{guo}: $H=\frac{3-\alpha}{2}$, which gives similar result
as calculated from the variance time plot in Fig.~\ref{fig-aggtimemod}.

%\enlargethispage{22pt}

In the following we look for the origin of the long range dependence
found above. In our model the topology and the TCP flow generating
rules are chosen such a way, that the link/source conditions for
self-similarity\cite{crovella,leland}  are excluded. In network-side models
the adaptive nature of TCP, namely the TCP backoff state mechanism is
pointed out as a main origin of such behavior\cite{fekete,guo}.
We investigate now if there is a close relation between the
self-similarity of the traffic, and backing off of the TCP.

In the backoff phase TCP retransmits the lost packet and doubles the
RTO.  TCP keeps track of the doubling times by a backoff variable $b$.
In the non-backoff phases $b=0$, and in backoff $b$ shows how many
times the RTO counter has been doubled. Due to Karn's algorithm the RTO is
doubled until two consecutive packet receives its acknowledgement.

First we recall shortly, that a TCP flow in the backoff phase produces
heavy-tailed statistics in the packet inter arrival time\cite{guo}. A TCP in a
given backoff state waits for a period $t=2^bT_{RTT}$ between two
packet sending attempts. The $b$-th backoff state occurs only after
$b$ successive packet losses. Let's denote the packet retransmission
probability (effective loss) with $p$. The probability of $b$
consecutive packet retransmission is $p^b$. Hence the probability of a
silent period $t$ due to backoffs, decays as $p^{log_2(t/T_{RTT})}\sim
t^{1-\alpha}$, where $\alpha=log_2(1/2p)$.

Next we repeat the main idea of a Markovian chain model for backoff
states\cite{fekete,guo} and show, that the statistics of backoff states
delivers the average effective loss probability $p$.

\begin{figure}[h]
\centerline{\resizebox{\kepnagysag}{!}{\includegraphics{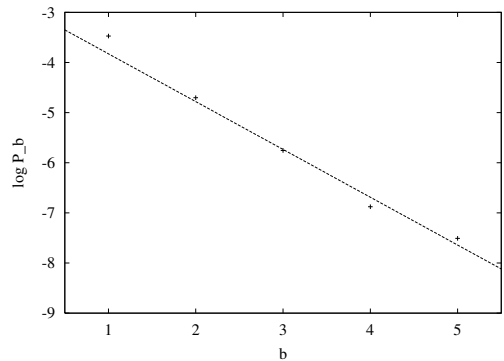}}}
\caption{Logarithmic plot of the probability distribution of backoff states 
($P_b$) as the function of the backoff variable $b$ The fitted line indicates
$p=21\%$ loss probability which gives a Hurst parameter $H=0.89$. 
\label{fig-backstatmod}}
\end{figure}

Let denote the probability $P_b$ that the TCP is in a $b$ deep
backoff. In a simplified Markovian framework one can estimate the
$P_b$ by the transition probabilities between backoff states as
follows (for a detailed matrix representation see \cite{figueiredo}.
The RTO value is doubled if one of two successive packets do not
receive ACK and is retransmitted. If the retransmission probability is
$p$ the transition probability to a deeper backoff is
$1-(1-p)^2=2p-p^2$.  This yields a backoff probability decay to be
$P_b\sim(2p-p^2)^b$ and one can read off the average loss probability
from the gradient of the semilogarithmic plot of $P_b$ versus $b$.  We
emphasize here, that the loss probability measured by the probability
of backoff states is the effective loss felt by the TCP. This
probability can be much larger as the real loss. This is the case in
our model, since the real loss is below $0.1\%$, however, the effective
loss is about $21\%$. A typical backoff distribution for our stochastic
model is shown in Fig.~\ref{fig-backstatmod}.

This gives us the possibility to demonstrate the connection between
long range dependency and the backoff distribution.  One compares the
probability $p$ calculated from the backoff statistics and the inter
packet arrival time decay factor $\alpha$ calculated from the packet
traffic time series. The two value agrees as $\alpha\approx log_2(1/2p)$, hence
the long range dependency is caused mainly by the backoff mode of the
TCP (and not by other external reasons as e.g.  long range distributed
user interaction).

%subsec

We have demonstrated the connection between the
long range dependency, self-similarity and backoff mechanism. Finally we
search for the origins of backing off the TCP.

%- results: no real packet loss, self similarity in flow and backoff\\

Our model by construction excludes the known origins of
self-similarity: the TCP flows follow each other immediately and
transfer data at a given rate without external silent
periods as e.g. would be the case with user-generated interactive
flow. The transfered file sizes are constant. The buffer lengths are
much above the critical chaotic threshold value\cite{fekete}. The only
stochasticity is in the switching between the destinations of the TCP
flows. This irregularity introduces some unpredictability in the data
flow. If this unpredictability is high, the TCP estimation for the
available sending rate is false. The consequences of unpredictability
has been studied from many aspects\cite{crovella,veresboda}, however all the 
previous studies require a
case when the high real loss probability (due to small buffers or
external loss) hinders the TCP to make sufficient predictions.

Here we presented a model, where the stochastic choosing of
destination itself pushes TCP into the backoff phase and generates
self-similar traffic. How can this happen?

TCP operates in backoff, if the ACK packet arrive after the limit set
by RTO. The RTO value is calculated from the traffic continuously,
using an average over some time-window. If the traffic is bursty,
with silent periods comparable with size of the averaging window, the TCP
cannot adapt to the rapid changes in the traffic. In our model we
detect heavy bursts in the queue lengths in the router. Since TCPs
changes the destination randomly, it can happen, that after a silent
period a buffer will be fed by one or two TCP. If these TCPs are in
slow start, the feeding of a buffer can be exponential fast.  The
queue lengths can hence grow very rapidly. If a queue gets longer,
packets arriving in this queue must wait longer. A rapid change in the
queue length can cause a so rapid change in the round trip time of a
packet, that the ACK for this packet arrives after the RTO expires.
So large fluctuations in the queue length (background traffic) can
cause a series of virtual losses and backing off the TCP. 

\begin{figure}[h]
\centerline{\resizebox{\kepnagysag}{!}{\includegraphics{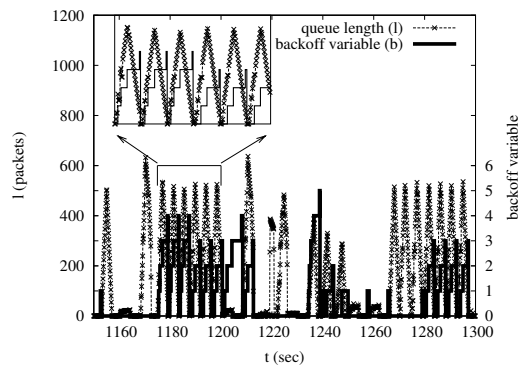}}}
%\resizebox{8cm}{!}{\includegraphics{empty.eps}}
\caption{Queue length in one of the buffers in the router $R$ and the backoff
  variable of the TCPs in a host. The time intervals where the queue
  length changes drastically overlap clearly with the backoff phases of
  the TCP. The inset shows a magnification of part of the data.
\label{fig-queuetime}}
\end{figure}

In Fig.~\ref{fig-queuetime} we show a typical queue length -- time
plot, where the large fluctuations cause backoff phase in a TCP. There
is a clear relation between the increasing queue length and the
evolution of backoff states.

% plot [550:650] 'que.10' w l 2,'tcp.2bck' u 2:(log($5)/log(2.))*100 w st 1 
  
Since in our model only the heavily fluctuating background traffic can
back off a TCP, we can conclude to identify the fluctuating background
as a source of self-similarity. This self-similarity is a
self-generated one, originating from the improper synchronization of
hosts, which continuously send data to each other by using many
successive TCP flows.

%\resizebox{7cm}{!}{\includegraphics{pk2.eps}}
%$\langle(\delta X^{m})^2\rangle\sim m^{2H}$
%
%H=0.85

\section{Case study: measurement}

In this section we present active measurement results which show
similar results in a real network environment as found in the previous
section in a small model scenario. The time evolution of a long TCP
flow on a transcontinental connection was followed on the IP level by
tcpdump\cite{tcpdump} and on the kernel level by a slightly modified
linux kernel from the 2.2.x series.  The modified kernel gave us the
possibility to follow directly the internal TCP variables {\em in
  situ} for a real network environment.

\begin{figure}[h]
\centerline{\resizebox{\kepnagysag}{!}{\rotatebox{270}{\includegraphics{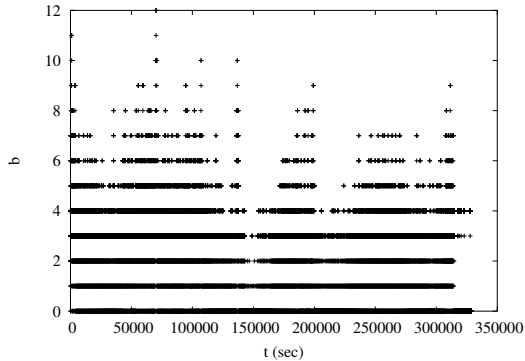}}}}
\caption{Backoff states of a TCP, which sends data through a congested 
transcontinental line.\label{fig-backtimemea}}
\end{figure}

On the transcontinental line chosen for the measurement typically many
TCP connections share the same route giving a highly fluctuating
background traffic. Additionally on the long line with many routers it
is possible that the packets of our TCP flow stacks in filled queues.
So the round trip time can fluctuate in a very wide range resulting
many backoff states. Figure~\ref{fig-backtimemea} shows a very
congested time interval, where many backoff states were observed. Here
we mention, that in contrast to the TCP implementations of ns-2, the
backoff variable $b$ of the linux kernel can have larger values than
6.

%- modified linux kernel

As described in the previous section the self-similarity is
characterized by the Hurst parameter, if the stochastic process under
consideration is weakly stationary. To satisfy this condition we
restrict our analysis only for some parts (time intervals) of the
whole measurement. 
%Later we show a more visual definition for these time intervals.

\begin{figure}[h]
\centerline{\resizebox{\kepnagysag}{!}{\includegraphics{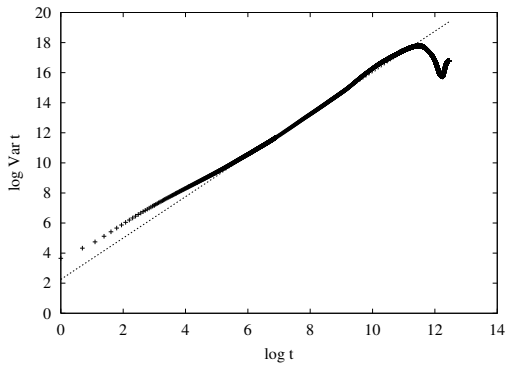}}}
\caption{Variance time plot of measured data in the longest stationary time 
interval. The fitted line indicates $H=0.69$.
\label{fig-vartimmea}}
\end{figure}

\begin{figure}[h]
\centerline{\resizebox{\kepnagysag}{!}{\rotatebox{270}{\includegraphics{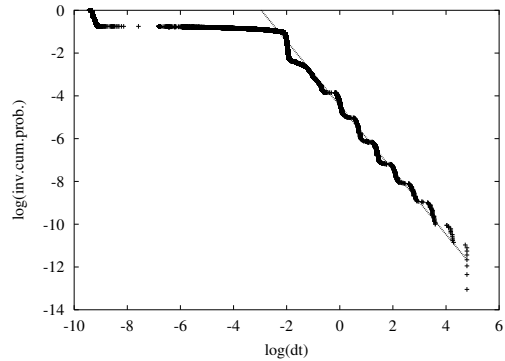}} }}
\caption{Statistics of inter arrival times. The fitted line on the slow 
decaying tail with $\alpha=1.505$ indicates long range dependent traffic. 
\label{fig-statintmea}}
\end{figure}

In the time range under study the highly congested traffic showed
self-similar nature. The variance time plot for the aggregated time
series of packet arrivals is plotted in Figure~\ref{fig-vartimmea},
from which we can read off the Hurst parameter $0.69$.  In
Fig.~\ref{fig-statintmea} we show the statistical distribution of
packet inter arrival times, which show an $\alpha=1.505$ decay giving a
similar value for the Hurst parameter as calculated from the variance
time plot.

%- congested transcontinental line: packet arrival and backoff statistics

Since we do not have total control over the whole internet, we cannot
prove rigorously that the observed self-similarity is the consequence
exclusively of the fluctuations in the background traffic as it is in
the simulation scenario presented in the previous section.  However it
is possible to show, that --- as in the simulation --- there is a
close relation between the inter packet time statistics and the
backoff statistics under such conditions where the real packet loss is
low, indicating self-generated self-similarity.

%subsec

Here we investigate first, what was the loss rate at the line. In
end-to-end measurements packet loss can be easily detected by
analyzing tcpdump data. But to gain this direct information about the
traffic, one needs special rights on the origin of the TCP flow and on
the destination as well. This ideal condition is given usually only
for a very restricted number of routes. In most cases one can monitor
the network traffic only on one side as it was the case in our
measurement. We applied the algorithm of Benko et.al.\cite{Benko} with some
improvement to detect packet losses from tcpdump data, and to decide
if the packet is lost really or timeout occurred.

The algorithm is the following. An effective loss occurs, if a packet is
resent.  A resent packet begins with the same serial number as the
original packet, so we have to count the number of packets, whose
sequence number occurred more than once in the TCP flow. We used
timestamps to avoid the wrapped sequence number problem.

Detecting real loss events is a bit more tricky. A sent packet is
probably lost if the TCP receives duplicate acknowledgement. Duplicate
ACKs are sent by the receiving TCP if a packet with higher sequence
number has arrived. However this can happen due to changes in the packet
order also, therefore the TCP waits a little and retransmits the
packet only if packet order switching is very improbable. Hence for
detecting real loss events we have to count the number of resent
packets, which are sent due to receiving of duplicate ACKs.

%In Table~\ref{tab-tcpdump}  we present a sort example from the tcpdump log 
%(irrelevant informations removed), which demonstrates, that detecting the
%{\em real} packet loss rate suffers from some uncertainty. In this
%example one cannot decide unambiguously if packets with numbers 46107
%and 47567 are lost or not.
%
%\begin{table}
%\caption{Part of a {\em tcpdump} log with irrelevant information removed.
%It cannot be decided unambiguously if packets 46107 and 47567 are lost or not.
%\label{tab-tcpdump}}
%{\tt
%\noindent
%5.4 IP.s > IP.r: P 41727:43187(1460) ack 1  \\
%5.5 IP.r > IP.s: . ack 40267 \\
%5.5 IP.s > IP.r: P 43187:44647(1460) ack 1  \\
%5.6 IP.r > IP.s: . ack 43187 \\
%5.6 IP.s > IP.r: P 44647:46107(1460) ack 1  \\
%5.6 IP.s > IP.r: P 46107:47567(1460) ack 1  \\
%5.7 IP.r > IP.s: . ack 44647 \\
%5.7 IP.s > IP.r: P 47567:49027(1460) ack 1  \\
%5.7 IP.s > IP.r: P 49027:50487(1460) ack 1  \\
%5.8 IP.r > IP.s: . ack 44647 \\
%5.8 IP.r > IP.s: . ack 44647 \\
%6.1 IP.s > IP.r: P 44647:46107(1460) ack 1  \\
%6.2 IP.r > IP.s: . ack 46107 \\
%6.2 IP.s > IP.r: P 46107:47567(1460) ack 1  \\
%6.2 IP.s > IP.r: P 47567:49027(1460) ack 1  \\
%7.2 IP.s > IP.r: P 46107:47567(1460) ack 1  \\
%9.1 IP.s > IP.r: P 46107:47567(1460) ack 1  \\
%9.3 IP.r > IP.s: . ack 50487 \\
%}
%\end{table}
%
%However we have carefully analyzed our data and found, that such ambiguous
%scenarios occur very rare giving just a small error in our estimation
%of real loss probability.

%subsec

\begin{figure}[h]
\centerline{\resizebox{\kepnagysag}{!}{\includegraphics{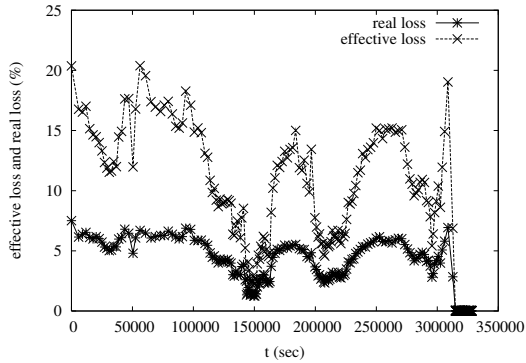}}}
\caption{Real loss and effective loss during the measurement. The loss rate
changes in time, therefore one has to restrict the analysis for the 
weakly stationary intervals.
\label{fig-lossmea}}
\end{figure}

Previously we mentioned that the background traffic during the whole
measurement cannot be approximated by weakly stationary stochastic
processes and for analysis one has to consider only parts of the data.
In this parts the flow can be characterized by static parameters e.g.
the loss ratio is constant in time. These intervals cannot be too
short to have enough data for statistical considerations. In
Fig.~\ref{fig-lossmea} we plot the loss probability versus time for
the whole measurement.  One can see long plateaus however there are
non-stationary regimes as well.  In the following we restrict ourself
only for the longest stationary range.

\begin{figure}[h]
\centerline{\resizebox{\kepnagysag}{!}{\includegraphics{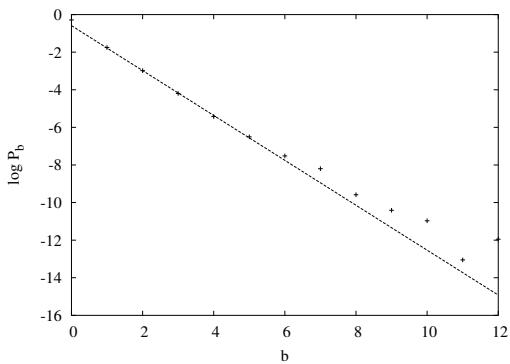}}}
%\rotatebox{270}{\resizebox{7cm}{!}{\includegraphics{backloss.ps}}}
\caption{Logarithmic plot of the backoff value probabilities 
observed in a transcontinental TCP data flow. The fitted line
indicates a $p=16.5\%$ loss rate felt by the TCP.
\label{fig-backstatmea}}
\end{figure}

We investigated the statistics of the backoff states for this time
regime from the data logged by the modified linux kernel.  We found,
that the distribution shows an exponential decay as it follows from
the Markovian description presented in the previous section. The
Fig.~\ref{fig-backstatmea} shows the decay of the probability of the
backoff states. The slope of the fitted line indicates a loss
probability $p=16.5\%$ felt by the TCP. This loss rate is consistent with
the asymptotic decay of the packet inter-arrival times
(Fig.~\ref{fig-statintmea}) and with the Hurst parameter of the
aggregated traffic (Fig.~\ref{fig-vartimmea}).

\begin{figure}[h]
\centerline{\resizebox{\kepnagysag}{!}{\includegraphics{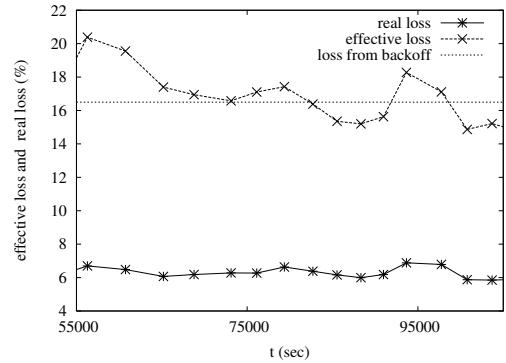}}}
\caption{Effective loss and real loss rate for a time interval. The horizontal
line shows the loss probability calculated from the backoff states.
\label{fig-lossbackmea}}
\end{figure}

So the close relation between the backoff states and the
self-similarity of the traffic holds. The next question is, if the TCP
is backed off due to real packet losses or the packets where only
delayed and timed out. In Fig.~\ref{fig-lossbackmea} we compare the
loss ratio from the backoff statistics ($16.5\%$) with the loss
probability calculated from the tcpdump output. We find, that the
average loss probability felt by the TCP equals with the {\em real
  plus virtual} (effective) loss and not with the {\em real} loss
alone. Here the difference between the two type of losses is crucial,
since the real loss is smaller than $12.5\%$, the lower bound of loss
probability, which gives self-similar traffic, but the effective loss
is higher.

%- real and effective loss: our estimation method

%- results: loss which cause the self-similarity is the effective loss and not
%the real packet loss

\section{Conclusion}

We have demonstrated in a model network and in a real measurement how
TCP can generate self-similar traffic itself. We showed that at very
low packet loss rate the congestion control mechanism detects false
packet losses if the background traffic is bursty. On this fluctuating 
background traffic TCP responds with backoff states. The switching between
backoff and congestion avoidance phases introduces further fluctuations
into the overall traffic, which results additional backoffs.
This self-generated burstiness evolves to self-similarity
however the network properties indicate simple, easily predictable traffic.

\enlargethispage{12pt}
In the future we focus on the self-generation of the burstiness,
what are the exact conditions for emergence of self-similarity in perfect
network.

%- discussion: self-generating burstiness, heavy tails, self similarity
%burstiness: short file transfers are in slow start, every slow start
%adds a burst to the background traffic.

\section*{Acknowledgements}

G. V. thanks the support of the Hungarian Science Fund {\nobreakspace}OTKA T037903 and
T 032437. P. P. thanks the support of the Hungarian Science Fund {\nobreakspace}OTKA D37788, T34832, T42981.

%\end{references}


\begin{thebibliography}{00}
%\begin{references}

\bibitem{Benko} P.Benko and A.Veres, ``A Passive Method for Estimating
  End-to-End TCP Packet Loss'', Globecom 2002, Taipei

\bibitem{csabai} I.Csabai, ``1/f noise in computer network traffic'',
  J. Phys. A: Math. Gen.,27, L417-L421, 1994.

\bibitem{crovella} M.Crovella and A.Bestavros, ``Self-similarity in
  world wide web traffic: Evidence and possible causes'', IEEE/ACM
  Transactions on Networking, pp. 835-846, December 1997.
  
\bibitem{danzig} P.Danzig, S.Jamin, R. C\'aceres, D.Mitzel and D.Estrin,
  ``An empirical workload model for driving wide-area TCP/IP network
  simulations'', Internetworking: Research and Experience 3:1-26, 1992.

\bibitem{erramilli} A.Erramilli, R.P. Singh and P. Pruthi, ``Chaotic
  Maps As Modesl of Packet Traffic'', Proc. of the 14th ITC, 329-338,
  June, 1994.

\bibitem{fekete} A.Fekete and G.Vattay, ``Self-Similarity in
  Bottleneck Buffers'' Proceedings of Globecom, 2001

\bibitem{feldmann} A.Feldmann, A.C.Gilbert, W.Willinger and T.G.Kurtz,
  ``The changing nature of network traffic: Scaling phenomena'', ACM
  Computer Communication Review, pp. 5-29, April 1998.
  
\bibitem{figueiredo} D.R.Figueiredo, B.Liu, V.Mishra and D.Towsley,
  ``On the autocorrelation structure of TCP traffic'', Tech. Rep.
  00-55, Dep. of Computer Science, University of Massachusetts,
  Amherst, November 2000.
  
\bibitem{guo} L.Guo, M.Crovella and I.Matta, ``TCP congestion control
  and heavy tails'', Tech. Rep. BUCS-TR-2000-017, Computer Science
  Dep., Boston University, 2000.

\bibitem{jacobson88} V.Jacobson, ``Congestion avoidance and control,'' in ACM
SIGCOMM,  314, 1998.

\bibitem{jacobson90} V.Jacobson, ``Modified TCP congestion avoidance
algorithm,'' Tech. Rep., end2end-interest mailing list, April 1990.

\bibitem{karn} P.Karn and C.Partridge, ``Improving round-trip time
  estimates in reliable transport protocols'', ACM Transactions on
  Computer Systems (TOCS), vol. 9, pp 365-373, 1991.

\bibitem{leland} W.E.Leland, M.S.Taqqu, W.Willinger and D.V. Wilson,
  ``On the self-similar nature of ethernet traffic (extended
  version)'', IEEE/ACM Transactions on Networking, pp. 1-15, February
  1994.

\bibitem{netsim} UCB, LBNL and VINT, ``Network simulator - ns (version 2)'',
http://www-mash.cs.berkeley.edu/ns, 2002.
  
\bibitem{park} K.Park, G.Kim and M.Crovella, ``On the
relationship between file sizes, transport protocols, and self-similar
network traffic'', In Proceedings of the International Conference on Network
Protocols, pp. 171-180, Oktober 1996


\bibitem{paxsonfloyd} V.Paxson and S.Floyd, ``Wide-area traffic: the
  failure of Poisson modeling. IEEE/ACM Transactions on Networking
  3:226-244, 1995.

\bibitem{tcpdump} LBL tcpdump, http://www.tcpdump.org.

\bibitem{veresboda} A.Veres and M.Boda, ``The chaotic nature of TCP
  congestion control'', in IEEE INFOCOM 2000, March 2000.

\bibitem{veresvattay} A. Veres, Zs. Kenesi, S. Molnar, and G. Vattay, 
``On the Propagation of Long-Range Dependence in the Internet,''
Computer Communication Review 30, No 4, pp. 243-254, 2000


\end{thebibliography}
\end{document}